\documentclass[conference]{IEEEtran}
\IEEEoverridecommandlockouts
\usepackage{cite}
\usepackage[numbers,sort&compress]{natbib}
\usepackage{amsmath,amssymb,amsfonts}
\usepackage{algorithmic}
\usepackage{graphicx}
\usepackage{textcomp}
\usepackage{xcolor}
 \usepackage{bbding}
 \usepackage{hyperref}
 \usepackage{authblk}

\hypersetup{
    colorlinks=true,  
    linkcolor=red,  
    filecolor=magenta,  
    urlcolor=red,  
    citecolor=blue,
}

\def\BibTeX{{\rm B\kern-.05em{\sc i\kern-.025em b}\kern-.08em
    T\kern-.1667em\lower.7ex\hbox{E}\kern-.125emX}}
\begin{document}

\title{Test Time Training for 4D Medical Image Interpolation\\
{\footnotesize \textsuperscript{}}
}

\author{
    Qikang Zhang\textsuperscript{1*}\thanks{*Correspondence to}, Yingjie Lei\textsuperscript{1}, Zihao Zheng\textsuperscript{2}, Ziyang Chen\textsuperscript{2}, Zhonghao Xie\textsuperscript{2}
}
\affil{
    \textsuperscript{1,2}Aberdeen Institution of Data Science and Artificial Intelligence, South China Normal University, Foshan, China \\
}

\maketitle

\begin{abstract}
4D medical image interpolation is essential for improving temporal resolution and diagnostic precision in clinical applications. Previous works ignore the problem of distribution shifts, resulting in poor generalization under different distribution. A natural solution would be to adapt the model to a new test distribution, but this cannot be done if the test input comes without a ground truth label. In this paper, we propose a novel test time training framework which uses self-supervision to adapt the model to a new distribution without requiring any labels. Indeed, before performing frame interpolation on each test video, the model is trained on the same instance using a self-supervised task, such as rotation prediction or image reconstruction. We conduct experiments on two publicly available 4D medical image interpolation datasets, Cardiac and 4D-Lung. The experimental results show that the proposed method achieves significant performance across various evaluation metrics on both datasets. It achieves higher peak signal-to-noise ratio values, 33.73dB on Cardiac and 34.02dB on 4D-Lung. Our method not only advances 4D medical image interpolation but also provides a template for domain adaptation in other fields such as image segmentation and image registration. The code is available at \href{https://github.com/ChaosTheProducer/TTT4D}{\textit{TTT4DMII}}.

\end{abstract}

\begin{IEEEkeywords}
4D medical image interpolation, test time training, self-supervised learning, domain adaptation
\end{IEEEkeywords}

\section{Introduction}
4D medical image interpolation focuses on generating intermediate frames from a sequence of medical images, helping to create smoother and more detailed representations of organs or tissues over time. This technique is especially useful in capturing subtle changes during procedures like heartbeats or lung movement, which are critical for accurate diagnosis and treatment. While video interpolation methods are widely applied in fields like film and animation, the unique constraints and requirements of medical imaging make it challenging to apply these methods to 4D medical image interpolation.

In CT scans, patients are exposed to higher levels of radiation, which can increase the risk of secondary cancers, making data collection challenging. In MRI, the data acquisition rate is slow, leading to motion artifacts such as blurring caused by factors like patient movement, unstable breathing, and difficulty maintaining a steady position during long scan times.

To tackle these challenges, many state-of-the-art methods have been proposed. Some of them focus on capturing periodic organ motion by improving model structure \cite{Guo2020ASV, Wei2023MPVF4M} or leveraging Transformer architectures to address the limitations of CNNs in modeling long-range spatial dependencies \cite{CHEN2022102615}, whereas others put their attention on optimization methods to enhance training and inference efficiency \cite{jia2023fouriernetleveragingbandlimitedrepresentation, JOSHI2023102917}. In addition, some researchers propose novel auxiliary losses to better improve the medical image interpolation task \cite{pmlr-v172-wolterink22a}.

Despite great progress previous works have made, very few of them care about distribution shifts. For example, during deployment, due to different imaging devices, the data received by the model and the data used for training do not come from the same distribution. This discrepancy often leads to poor performance, which inspires us to borrow the idea from Test-Time Training (TTT) and propose a TTT-based training paradigm. 

\begin{figure*}
    \centering
    \includegraphics[width=1\linewidth]{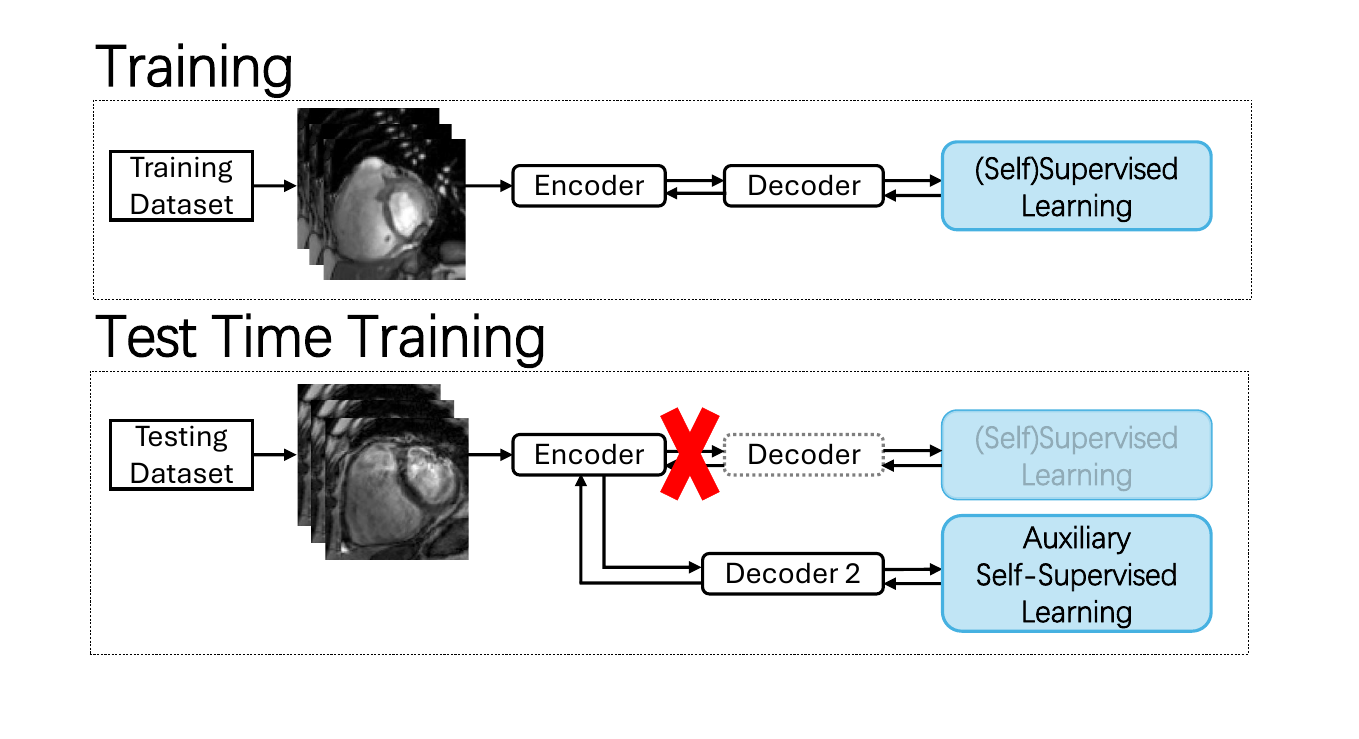}
    \caption{\textbf{Up}: Training before deployment. A standard training workflow, we optimize the self-supervised objective. \textbf{Down}: Training during deployment. Observations are collected from the test set, and we optimize only the auxiliary self-supervised objective.}
    \label{fig:teaser}
\end{figure*}

The basic idea of TTT is to use self-supervised learning to adapt the model to a new distribution \cite{sun2020test, sun2019unsupervised}. We modify the model during test time to enhance its performance on specific instance. The only issue is that the test data lacks the corresponding labels. But we can use self-supervised learning to generate pseudo-labels directly from the input data itself. As shown in figure \ref{fig:teaser}, the TTT framework resembles a sideways "Y" structure. The head represents the feature extractor, while the upper and lower branches correspond to the main task network and the self-supervised auxiliary network, respectively. During training, TTT optimizes both the main network and the self-supervised network. At test time, each test input is first processed through the upper branch to let the model adapt to the new distribution using the self-supervised auxiliary task, after which predictions are made for the main task.

\begin{figure*}
    \centering
    \includegraphics[width=1\linewidth]{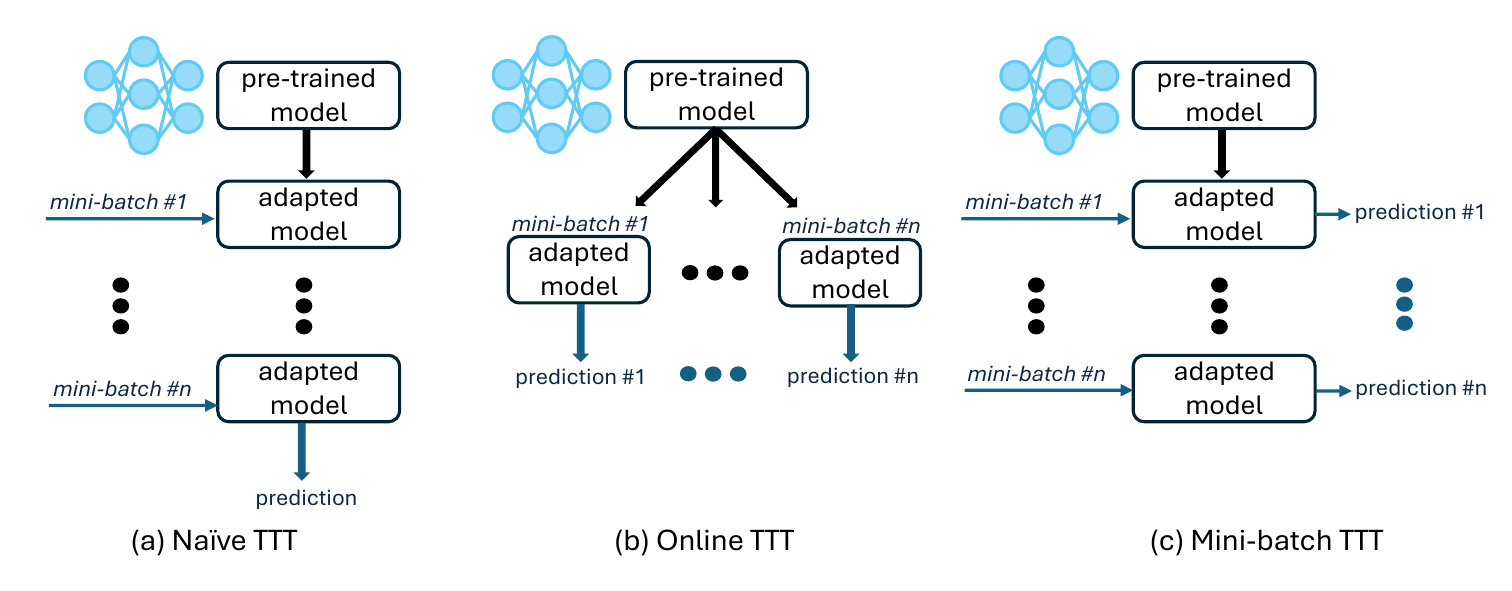}
    \caption{Visualization of three TTT schemes: (a) Naïve TTT, (b) Online TTT, (C) Mini-batch TTT. In (a) Naïve TTT, the model is adapted using all test samples before making predictions; (b) Online TTT, where the model is adapted individually for each mini-batch, with updates independent from other batches; and (c) Mini-batch TTT, where the model is adapted in an online manner to the entire test set, and previous knowledge can contribute to current one. }
    \label{fig:teaser2}
\end{figure*}

On top of this, we introduce the TTT design for 4D medical image interpolation(\textit{TTT4MII}) and incorporate two self-supervision tasks which are commonly used in computer vision: rotation prediction and image reconstruction. We consider the rotation prediction task a four-class classification task. To be specific, we randomly rotate the input (0°, 90°, 180°, or 270°) and pass it through the model to perform the classification task. For the image reconstruction task, inspired by masked autoencoders \cite{he2022masked}, we implement a 3D MAE model designed for slice-based image inputs. In this setting, we remove parts of the input image, and the model learns to predict the missing values, enabling it to adapt to a new distribution. Additionally, we explore three different TTT schemes(As shown in Figure \ref{fig:teaser2}): a) Naïve TTT, b) Online TTT, and c) Mini-batch TTT. Consider there are \( m \) unlabeled batches of test data \( \{ b_1, b_2, \dots, b_m \}\), and the pre-trained model starts with initial weights \( \theta_0 \). Firstly, in naïve TTT, the model adapts to all \( m \) batches through multiple epoch of optimization before making the final prediction. The final model weights \( \theta \) used for prediction are obtained after these adaptations. Secondly, in Online TTT, the pre-trained model adapts independently to each mini-batch. In this case, the weights \( \theta_i \) updated on one batch are independent of the weights \( \theta_j \) updated on another batch, as all adaptations start directly from  \( \theta_0 \). Thirdly, in Mini-batch TTT, the model sequentially adapts to the target dataset \( \{ b_1, \dots, b_m \} \) in a stream, where each mini-batch is observed only once. In other words, the model weights are updated iteratively, with each adaptation building upon the knowledge learned from the previously batches. Experiments show that our method can effectively adapt to unseen test distributions, achieving state-of-the-art performance on the Cardiac and 4D-Lung datasets. 

The main contributions of this paper are summarized as follows:
\begin{itemize}
    \item We are the first to introduce TTT in 4D medical image interpolation.
    \item We propose a novel test time training framework for 4D medical image interpolation, \textit{TTT4MII}, which enables model to adapt to the test distribution without any labels.
    \item We explored three different TTT schemes: a) Naïve TTT, b) Online TTT, and c) Mini-batch TTT. Moreover, we conduct experiments to analyze how they affect the performance of the main task.
    \item Experiments demonstrate that our method significantly improves interpolation accuracy and efficiency across multiple medical imaging benchmarks.
\end{itemize}

\section{Related Work}
\subsection{4D Medical Image Interpolation}
4D medical image interpolation tackles the challenge of generating high-resolution temporal data, which is often limited by factors such as radiation exposure and scan time. VoxelMorph provides a fast, learning-based framework for generating deformation fields, optimizing registration and interpolation tasks with convolutional networks \cite{balakrishnan2019voxelmorph}. Another approach by Kim introduced a diffusion deformable model, which integrates diffusion and deformation modules to generate intermediate frames along a geodesic path while preserving spatial topology \cite{Kim2022DiffusionDM}. Additionally, Kim proposed UVI-Net, an unsupervised framework that directly interpolates temporal volumes without intermediate frames, demonstrating robustness with minimal training data \cite{kim2024data}. However, a few of these methods consider the impact of domain shifts, which can significantly degrade model performance in practical clinical settings.

\subsection{Test Time Training}
TTT improves model adaptability to distribution shifts by updating parameters during inference \cite{liang2024comprehensive}. In image classification, Sun formulated TTT as a self-supervised task on test samples, achieving robust performance under domain shifts \cite{sun2020test, sun2019unsupervised, gandelsman2022test}. In anomaly detection and segmentation, Costanzino leveraged TTT to use features from test data for training a binary classifier, improving segmentation accuracy without labeled anomalies \cite{Costanzino2024TestTT}. In video object segmentation, Bertrand incorporated mask cycle consistency in TTT to counter performance drops caused by video corruptions and sim-to-real transitions \cite{bertrand2023testtime}.
   
\subsection{Self-supervised Learning}
Self-supervised learning leverages automatically generated labels from data itself to learn meaningful representations without requiring manual annotations. A common strategy in SSL is to design auxiliary tasks with specific loss functions that guide the model to extract relevant features \cite{noroozi2016unsupervised, chen2019self, taleb2021multimodal}. For instance, Doersch proposed a jigsaw puzzle task where an image is split into patches, and the network predicts their spatial arrangement, promoting an understanding of spatial structure \cite{doersch2015unsupervised}. Similarly, Gidaris used rotation prediction as an auxiliary task, where the model identifies the rotation angle (0°, 90°, 180°, or 270°) applied to an image, enhancing its sensitivity to geometric transformations \cite{gidaris2018unsupervised}. More recently, Chen introduced contrastive learning via SimCLR, which uses a contrastive loss to maximize agreement between augmented views of the same instance, learning invariant representations across transformations \cite{chen2020simple}. These approaches highlight the versatility of self-supervised learning in capturing robust data representations. 

\section{Methods}\label{Method}
In this section, we describe our proposed \textit{TTT4MII} method. It can be implemented on top of any generic model architecture, such as feature extractor-prediction head or encoder-decoder frameworks.
\subsection{Problem Setup}
Formally, given a video \( V \) consist of \( n \) frames, represented as \( \{I_0, I_1, \dots, I_{n-1}\} \), where each \( I_i \) corresponds to a 3D medical image at time \( T = \frac{i}{n-1} \). Given two consecutive frames \( I_0 \) and \( I_{n-1} \) at \( T = 0 \) and \( T = 1 \), respectively, our objective is to predict an intermediate frame \( \hat{I}_t \) at any temporal point \( T = t \), where \( 0 < t < 1 \). 

\subsection{Network Architecture}
Our architecture is designed to allow the interpolation network and the self-supervised auxiliary network to share features. The architecture includes a feature extractor \( f \), which is shared by both the main task head \( h \) and the self-supervised auxiliary head \( g \). To be specific, \( f \) is a 3D AlexNet, while \( h \) is UVI-NET, the interpolation head for predicting the next frame. We define \( f \) as the feature extractor with parameters \( f_\theta \) and \( h \) as the main task head with parameters \( h_\theta \), such that \( model(I; \theta) = h(f(I)) \), where \( I \) represents the input frames. Intuitively, our method aims to update \( f \) (and thus \( f_\theta \)) during test time using gradients from the auxiliary head \( g \), allowing \( f \) to adapt to the test distribution. The self-supervised auxiliary head \( g \) with parameter \( g_\theta \), takes the output of \( f \) as its input (as shown in Figure \ref{fig:teaser}). In this work, we employ two self-supervised tasks: rotation prediction and image reconstruction.

\subsection{Rotation Prediction and Masked Autoencoders}
We use rotation prediction as one of the self-supervised auxiliary tasks. To be specific, we rotate the input frame by one of four possible angles (0°, 90°, 180°, and 270°) and input this rotated frame to the model. The task is defined as a four-way classification problem, where the model predicts the rotation angle of the input image. During TTT, the auxiliary task generates pseudo-labels for rotation, and the model adapts to the test distribution by minimizing the cross-entropy loss:  

\[
    \mathcal{L}_g = -\frac{1}{N} \sum_{i=1}^N \sum_{c=1}^4 \theta_{i,c} \log(\hat{\theta}_{i,c})
\]

where \( \hat{\theta}_{i,c} \) is the predicted probability for the \( c \)-th rotation angle of the \( i \)-th input frame, \( \theta_{i,c} \) is the ground truth label for self-supervision, and \( N \) is the number of input frames. This auxiliary task allows the model to learn spatial and structural representations beneficial for the interpolation process, helping the model adapt to unseen data by updating its parameters.

As an alternative self-supervised task, we use image reconstruction. In image reconstruction, a simple yet effective method is MAE, which masks a large proportion of patches to create a challenging task, helping the model learn generalizable features. We extend the original MAE architecture by replacing the ViT with 3D-ViT so that it can handle 3D inputs. Each input image \( x \) is first divided into many small patches. Then, we randomly mask 80\% of the patches in \( x \) and feed the remaining patches into an autoencoder. The self-supervised objective \( L_g(g \circ f(x), x) \) compares the masked patches reconstructed by \( g \circ f(x) \) with the original masked patches in \( x \) and compute the pixel-wise mean squared error.

\subsection{Test Time Training}
Here, we introduce three different schemes of TTT: a) Naïve TTT, b) Online TTT, and c) Mini-batch TTT. The objective for all three is to optimize the loss function:

\[
\min \mathbb{E} = \mathcal{L}_g(g \circ f(x), x)
\]

\textbf{Naïve TTT}: In this setting, we adapt the model using all the test data batches at once before making predictions. This scheme assumes access to a collection of unlabeled batches \( b_1, b_2, \dots, b_m \) and the pre-trained model weights \( \theta_0 \). The model performs multiple epoch across all batches to optimize the weights, producing an optimized model \( (f\theta, g\theta) \). This naive idea is that the model adapts globally access all visible data, without considering the independence of each individual test batch. This setting allows the model to adjust its weights to better match the target distribution:

\[
\theta = \arg \min_{\theta} \sum_{i=1}^{m} \mathcal{L}_g(g \circ f(b_i), b_i)
\]

\textbf{Online TTT}: Here we adapt the model to each test batch independently. That is, the weights \( \theta_i \) updated by the model on batch \( b_i \) are independent from those updated on other batches. In this setting, the model updates independently as new test batch arrives, and each batch adaptation is independent of the others. Formally, for each batch \( b_i \), the model is updated as follows:

\[
\theta_i = \theta_0 - \eta \nabla_{\theta} \mathcal{L}_g(g \circ f(b_i), b_i)
\]

where \( \eta \) is the learning rate, and all \( \theta_i \) updates directly from the initial weights \( \theta_0 \).

\textbf{Mini-Batch TTT}: In Mini-batch TTT, we adapt the model in an online manner but with a slight modification: we update the model incrementally across all batches. Here, the main idea is that knowledge learned from previous batches can help the model adapt better to batches which come later. This incremental learning process ensures that the model retains and improves upon what it has learned from earlier data. To be specific:

\[
\theta_i = \theta_{i-1} - \eta \nabla_{\theta} \mathcal{L}_g(g \circ f(b_i), b_i)
\]

\section{Experimental Settings}
\subsection{Datasets}
We conduct our experiments on the Cardiac and the 4D-Lung dataset.

The Cardiac dataset includes 100 heart scans, capturing motion between the end-diastolic and end-systolic phases. The dataset is split into 90 scans for training and 10 scans for testing. On average, there are 10 frames between these two phases, providing the temporal resolution necessary for evaluating interpolation methods.

The 4D-Lung dataset consists of 82 chest CT scans from 20 lung cancer patients, taken at the end-inspiratory and end-expiratory phases. The training data comprises scans from 18 patients, while the remaining 2 are used for testing. The dataset is preprocessed to highlight lung-specific features, with normalization, windowing, and bed removal. All images are resized to 128×128×128 for consistency.

\subsection{Evaluation Metrics}
To evaluate the performance of our model, we use five common metrics for image interpolation: \textbf{PSNR} (Peak Signal-to-Noise Ratio), \textbf{NCC} (Normalized Cross-Correlation), \textbf{SSIM} (Structural Similarity Index), and \textbf{NMSE} (Normalized Mean Squared Error). PSNR measures the quality of the reconstructed images by comparing their pixel-wise differences. \textbf{NCC} assesses the similarity between the predicted and ground truth frames by measuring correlation. \textbf{SSIM} evaluates the structural similarity, considering luminance, contrast, and texture. \textbf{NMSE} quantifies the difference between the predicted and true images. The metrics we use can provide a comprehensive evaluation of both the quality and perceptual fidelity of interpolated images.

\subsection{Implementation Details}
Our experiments are conducted on an Ubuntu 22.04 environment using a NVIDIA RTX 4090 GPU. During training time, We train our models for 200 epochs with a learning rate of $ 2 \times 10^{-4} $ and 50 epochs for testing time. Using a batch size of 1 is not only necessary to fit the data within the available GPU memory but also more reflective of real-world scenarios, where models are often applied to single patient images or scans at a time.

The base model we use is UVI-Net \cite{kim2024data}, two U-Net architecture composed of a reconstruction model and an optical flow calculator. After the training phase, we freeze the parameters of the decoder during testing. To implement the auxiliary self-supervised task, we introduce a simple prediction head consisting of two fully connected layers to predict the rotation angle of input images. And we extend MAE to a 3D variant, so the encoder can adapt to the new distribution of unseen data. We randomly mask 80\% of the input patches and task the encoder with reconstructing the missing regions during training. The auxiliary task facilitates back propagation to update the encoder parameters. During TTT, the model adapts to unseen test data by leveraging this self-supervised optimization. Once TTT concludes, we proceed with interpolation and evaluate the results. This approach make the encoder adapt to the test distribution, improving the robustness and accuracy of the interpolation process. As shown in Figure \ref{fig:teaser}, the encoder is updated during test time using the auxiliary task of rotation prediction or 3D MAE, which enhances the model’s robustness and accuracy when processing unseen data.

We assess the impact of different self-supervised task and TTT scheme on interpolation task in Cardiac and 4D-Lung datasets. And We compare the results of our method with previous methods to show TTT provides a measurable improvement in accuracy and efficiency.  

\section{Experimental Results}
\subsection{Comparisons with Previous Methods}
As shown in table \ref{tab:cardiac} and table \ref{tab:4d-lung}, The experimental results demonstrate the effectiveness of our proposed method across two datasets: Cardiac and 4D-Lung. We compare our method with previous methods. Key evaluation metrics include PSNR, NCC, SSIM, and NMSE, with higher PSNR, NCC, and SSIM indicating better performance and lower NMSE indicating reduced error and perceptual dissimilarity.

On the Cardiac dataset, our method achieves the highest PSNR (33.73), NCC (0.571), and NMSE (2.230), surpassing the strongest baseline, UVI-Net, which achieves a PSNR of 33.59 and an NCC of 0.565. Our approach also records the lowest NMSE (2.230), reflecting improved interpolation accuracy and perceptual quality. Compared to the widely used VM model, our method delivers a significant improvement of over 2.7 dB in PSNR.

On the 4D-Lung dataset, our method similarly outperforms all baselines. It achieves the highest PSNR (34.02), SSIM (0.320), and NCC (0.981). Additionally, our method reduces NMSE to 0.551, further confirming its superior performance in handling lung motion. These results show notable improvements over UVI-Net, which achieves a PSNR of 34.00 and an SSIM of 0.980. Compared to VM, our method demonstrates a 1.7 dB improvement in PSNR.

The results show that our TTT framework with auxiliary loss not only improves interpolation accuracy but also generalizes well to unseen data, achieving consistent improvements over other state-of-the-art baselines. Our method is particularly effective in scenarios with domain-specific challenges, such as cardiac motion and lung deformation. These outcomes highlight the practical benefits of combining self-supervised learning and model efficiency optimization for medical video interpolation tasks. 

\begin{table}
    \centering
    \begin{tabular}{c|cccc}
         \hline
          Method&  PSNR$\uparrow$ &  NCC$\uparrow$&  SSIM$\uparrow$&  NMSE$\downarrow$\\
         \hline
           SVIN\cite{Guo2020ASV}&32.51&0.559&0.972&2.930\\
           MPVF\cite{Wei2023MPVF4M}&33.15&0.561&0.971&2.435\\
           VM\cite{balakrishnan2019voxelmorph}& 31.02 & 0.555 & 0.966 & 4.254\\
           TM\cite{CHEN2022102615}&30.45&0.547&0.958&4.826\\
           Fourier-Net+\cite{jia2023fouriernetleveragingbandlimitedrepresentation}&29.98&0.544&0.957&5.503\\
           R2Net\cite{JOSHI2023102917}&28.59&0.509&0.930&7.281\\
           DDM\cite{Kim2022DiffusionDM}& 29.71 & 0.541 & 0.956 & 5.007 \\
           IDIR\cite{pmlr-v172-wolterink22a}&31.56&0.557&0.968&3.806\\
           UVI-Net\cite{kim2024data}& 33.59 & 0.565 & \textbf{0.978} & 2.384\\ \hline
           \textbf{Ours}& \textbf{33.73} & \textbf{0.571} & \textbf{0.978} & \textbf{2.230}\\ \hline
    \end{tabular}
    \caption{Comparison of our model and other competitive models on Cardiac dataset. The $\uparrow$ indicates that the larger the value, the better the performance, while the $\downarrow$ indicates that the smaller the value, the better the performance}
    \label{tab:cardiac}
\end{table}

\begin{table}
    \centering
    \begin{tabular}{c|cccc}
         \hline
          Method&  PSNR$\uparrow$ &  NCC$\uparrow$&  SSIM$\uparrow$&  NMSE$\downarrow$\\
         \hline
           SVIN\cite{Guo2020ASV}&30.99&0.312&0.973&0.852\\
           MPVF\cite{Wei2023MPVF4M}&31.18&0.310&0.972&0.761\\
           VM\cite{balakrishnan2019voxelmorph}& 32.29 & 0.316 & 0.977 & 0.641\\
           TM\cite{CHEN2022102615}&30.92&0.313&0.973&0.786\\
           Fourier-Net+\cite{jia2023fouriernetleveragingbandlimitedrepresentation}&30.26&0.308&0.971&0.959\\
           R2Net\cite{JOSHI2023102917}&29.34&0.294&0.962&1.061\\
           DDM\cite{Kim2022DiffusionDM}& 30.27 & 0.308 & 0.971 & 0.905\\
           IDIR\cite{pmlr-v172-wolterink22a}&32.91&0.321&0.980&0.586\\
           UVI-Net\cite{kim2024data}& 34.00 & \textbf{0.320} & 0.980 & 0.552\\ \hline
           \textbf{Ours}& \textbf{34.02} & \textbf{0.320} & \textbf{0.981} & \textbf{0.551}\\ \hline
    \end{tabular}
    \caption{Comparison of our model and other competitive models on 4D-Lung dataset. The $\uparrow$ indicates that the larger the value, the better the performance, while the $\downarrow$ indicates that the smaller the value, the better the performance}
    \label{tab:4d-lung}
\end{table}

\begin{table}
    \centering
    \begin{tabular}{c|c|c}
    \hline
         Method&Scheme& Time Cost \\ \hline 
         Rotation Prediction&Naive TTT  &0.93766s per \\
         &Online TTT &0.91476s per \\
         &Mini-Batch TTT  &0.90164s per \\ \hline
         3D-MAE&Naive TTT  &2.9068s per \\
         &Online TTT  &2.43s per \\
         &Mini-Batch TTT  &2.3524s per \\ \hline
    \end{tabular}
    \caption{Comparison of inference times for different self-supervised tasks and schemes on Cardiac Dataset.}
    \label{tab:timecost}
\end{table}

\subsection{Ablation study}

\begin{table*}
    \centering
    \begin{tabular}{c|c|cc|cccc}
         \hline
          Dataset&Scheme&Rotation Prediction&3D-MAE&  PSNR$\uparrow$ &  NCC$\uparrow$&  SSIM$\uparrow$&  NMSE$\downarrow$\\
         \hline
         Cardiac&Naïve TTT& \checkmark &-&33.70 \tiny{$\pm{0.256}$} &0.568 \tiny{$\pm{0.011}$} &0.978 \tiny{$\pm{0.002}$} &2.263 \tiny{$\pm{0.282}$} \\
          &Online TTT& \checkmark &-&33.70 \tiny{$\pm{0.256}$} &0.568 \tiny{$\pm{0.011}$} &0.978 \tiny{$\pm{0.003}$} &2.263 \tiny{$\pm{0.281}$} \\
          &Mini-Batch TTT& \checkmark &-&33.70 \tiny{$\pm{0.256}$} &0.568 \tiny{$\pm{0.011}$} &0.978 \tiny{$\pm{0.002}$} &2.263 \tiny{$\pm{0.282}$} \\
          &NaïveTTT& - & \checkmark &\textbf{33.73} \tiny{$\pm{0.247}$} &\textbf{0.571} \tiny{$\pm{0.011}$} &0.978 \tiny{$\pm{0.002}$} &\textbf{2.230} \tiny{$\pm{0.273}$} \\
          &Online TTT& - & \checkmark &33.72 \tiny{$\pm{0.247}$} &\textbf{0.571} \tiny{$\pm{0.011}$} &0.978 \tiny{$\pm{0.002}$} &\textbf{2.230} \tiny{$\pm{0.274}$} \\
          &Mini-Batch TTT& - & \checkmark &\textbf{33.73} \tiny{$\pm{0.247}$} &\textbf{0.571} \tiny{$\pm{0.011}$} &0.978 \tiny{$\pm{0.002}$} &\textbf{2.230} \tiny{$\pm{0.272}$} \\
        \hline
         4D-Lung&Naïve TTT& \checkmark &-&33.98 \tiny{$\pm{0.336}$} &0.320 \tiny{$\pm{0.04}$} &0.980 \tiny{$\pm{0.003}$} &0.553 \tiny{$\pm{0.072}$}   \\
          &Online TTT& \checkmark &-&33.96 \tiny{$\pm{0.336}$} &0.320 \tiny{$\pm{0.04}$} &0.980 \tiny{$\pm{0.003}$} &0.554 \tiny{$\pm{0.072}$}   \\
          &Mini-Batch TTT& \checkmark & -&33.98 \tiny{$\pm{0.336}$} &0.320 \tiny{$\pm{0.04}$} &0.980 \tiny{$\pm{0.003}$} &0.553 \tiny{$\pm{0.072}$}   \\
          &NaïveTTT& - & \checkmark &\textbf{34.02} \tiny{$\pm{0.341}$} &0.320 \tiny{$\pm{0.04}$} &\textbf{0.981} \tiny{$\pm{0.003}$} &\textbf{0.551} \tiny{$\pm{0.077}$}  \\
          &Online TTT& - & \checkmark &\textbf{34.02} \tiny{$\pm{0.344}$} &0.320 \tiny{$\pm{0.04}$} &\textbf{0.981} \tiny{$\pm{0.002}$} &\textbf{0.551} \tiny{$\pm{0.077}$}   \\
          &Mini-Batch TTT& - & \checkmark &\textbf{34.02} \tiny{$\pm{0.344}$} &0.320 \tiny{$\pm{0.04}$} &\textbf{0.981} \tiny{$\pm{0.002}$} &\textbf{0.551} \tiny{$\pm{0.077}$} \\ \hline
         \hline
    \end{tabular}
    \caption{COMPARISON OF DIFFERENT SETTING ON Cardiac AND 4D-Lung DATASET. The $\uparrow$ indicates that the larger the value, the better the performance, while the $\downarrow$ indicates that the smaller the value, the better the performance}
    \label{tab:ttt}
\end{table*}

\textbf{Different Self-supervised Task}
We further analyze the impact of different self-supervised tasks on our TTT framework. In particular, we compare the performance of Rotation Prediction and 3D-MAE as auxiliary self-supervised learning tasks. The results are presented in Table \ref{tab:ttt}, where we evaluate their effects under three different TTT schemes: Naïve TTT, Online TTT, and Mini-Batch TTT.

As shown in Table \ref{tab:ttt}, the use of 3D-MAE consistently leads to better interpolation accuracy across all TTT schemes. For instance, on the Cardiac dataset, Naïve TTT with 3D-MAE achieves a PSNR of 33.73, which is higher than the 33.70 obtained with Rotation Prediction. Similarly, NCC improves from 0.568 to 0.571, and NMSE decreases from 2.263 to 2.230, indicating a more effective self-supervised signal when leveraging 3D-MAE.

A similar trend is observed in the 4D-Lung dataset, where 3D-MAE consistently provides a slight performance boost over Rotation Prediction. This suggests that the feature representations learned through 3D-MAE are more beneficial for adapting the model to unseen data, likely due to its ability to capture richer spatial-temporal information.

In addition to accuracy, we compare the inference time required for different self-supervised tasks, as shown in Table \ref{tab:timecost}. The results indicate that Rotation Prediction is computationally more efficient, with all three schemes taking approximately 0.9s per sample, whereas reconstruction task requires more than 2s per sample. Despite the inference time increases, the performance improves with the use of 3D-MAE.

A key finding is that there exists a relationship between task complexity and its effectiveness in TTT. Simple self-supervised tasks, such as rotation prediction, provide stable adaptation but offer limited performance gains. In contrast, more complex tasks, such as image reconstruction, yield greater benefits but may introduce instability in adaptation under certain conditions. This instability could stem from excessive gradient fluctuations caused by overly complex tasks, which in turn affect the model's convergence stability during testing. Moreover, we observe that different self-supervised tasks exhibit varying degrees of robustness to different types of distribution shifts. For instance, under specific domain shifts such as style variations, reconstruction-based tasks significantly outperform contrastive learning methods, whereas for shifts induced by geometric transformations, contrastive approaches prove more effective. This highlights the importance of selecting self-supervised tasks based on the specific nature of distribution shifts rather than relying on a fixed task across all scenarios.

\textbf{Different Scheme}
We observe that he performance differences among the three TTT schemes—Naïve TTT, Online TTT, and Mini-batch TTT—are surprisingly subtle, despite their distinct adaptation strategies. In Naïve TTT, the model undergoes multiple epochs of optimization over the entire test set before making predictions, allowing it to fully adapt to the target distribution. While this approach might seem advantageous, it risks overfitting to the test data, especially when the test set is small or lacks diversity. Online TTT, on the other hand, adapts independently to each test batch, starting from the same initial weights for every batch. This independence ensures robustness to distribution shifts within individual batches but may fail to leverage shared patterns across the test set. Mini-batch TTT strikes a balance by incrementally updating the model across batches, allowing knowledge from earlier batches to inform adaptations for later ones. This sequential adaptation mimics a form of continues learning, where the model continuously refines its understanding of the test distribution.

The minimal performance gap between these schemes can be attributed to several factors. First, the self-supervised tasks (rotation prediction and image reconstruction) are inherently designed to capture generalizable features, which reduces the sensitivity of the model to the specific adaptation strategy. Second, the shared feature extractor \( f \) ensures that the model retains a strong prior from the source domain, limiting the extent to which test-time updates can diverge. Finally, the relatively small size of the test batches in medical imaging scenarios may diminish the practical differences between the schemes, as the model’s adaptations are constrained by the limited data available at each step.

While the performance differences are small, the choice of scheme may still depend on practical considerations. Naïve TTT is suitable when the test set is large and diverse, allowing the model to benefit from global adaptation. Online TTT is ideal for scenarios where test data arrives sequentially, and computational efficiency is a priority. Mini-batch TTT offers a middle ground, providing incremental adaptation without the computational overhead of Naïve TTT. Ultimately, the robustness of the proposed framework lies in its flexibility to accommodate different adaptation strategies while maintaining strong performance across the board.

\begin{figure*}
    \centering
    \includegraphics[width=1\linewidth]{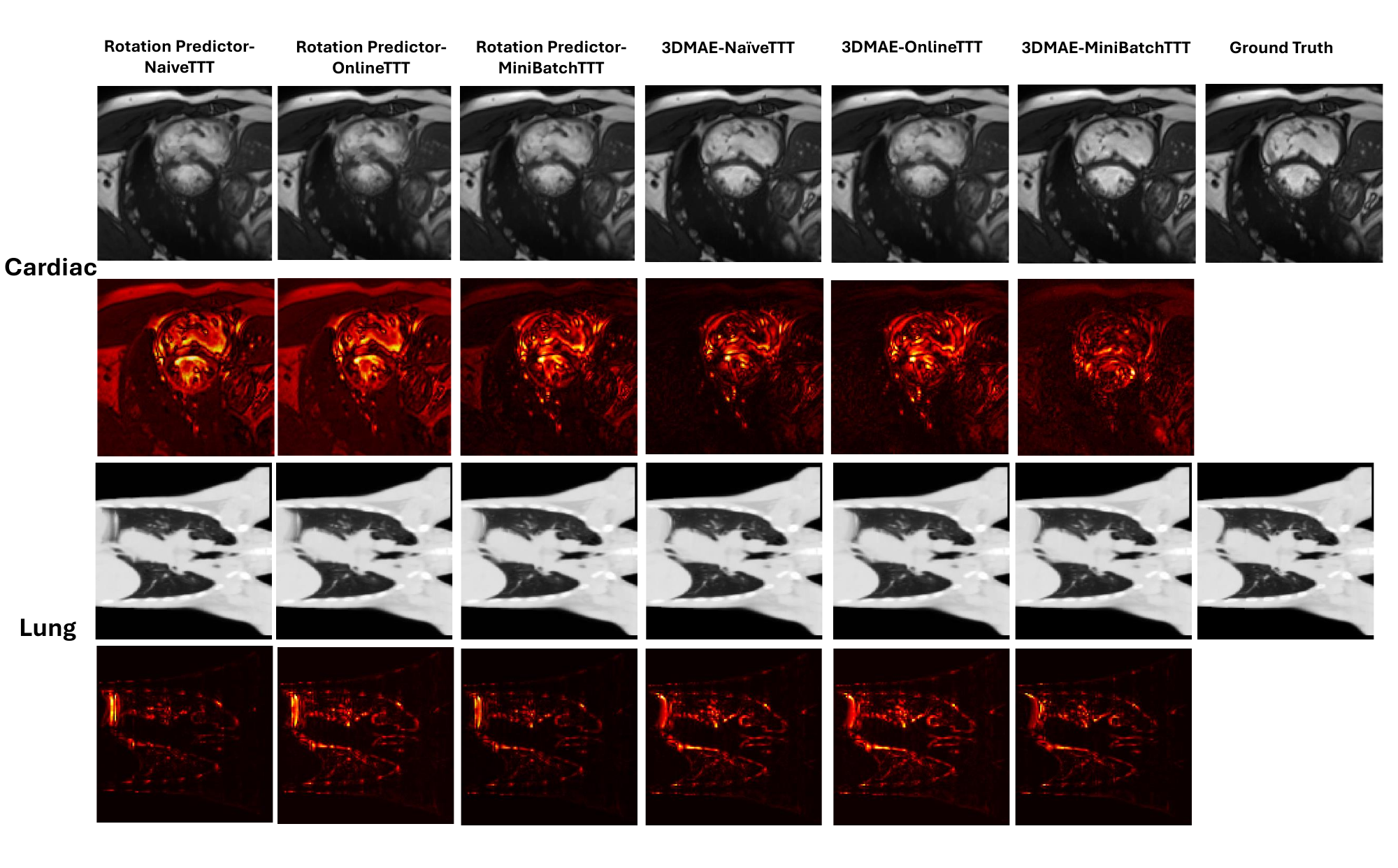}
    \caption{Visualization of interpolation results on the Cardiac and 4D-Lung dataset. The top row shows the predicted frames (left) and the ground truth (right) for six different settings, combining two self-supervised taskswith three TTT schemes. The bottom row visualizes the difference maps between the predictions and the ground truth, with warmer colors indicating larger errors.}
    \label{fig:lung}
\end{figure*}

\subsection{Qualitative Analysis}
We present a visualization to compare interpolation results in different Settings. Figure \ref{fig:lung} illustrates the interpolation results under six different settings, combining two self-supervised tasks with three TTT schemes. The top row displays the predicted frames alongside the ground truth, highlighting the overall structural similarity and temporal coherence of the interpolated results. The bottom row further extracts and visualizes the differences between the predictions and the ground truth, providing a detailed view of where each setting excels or falls short. The results reveal several key observations. First, both self-supervised tasks produce predictions that are visually close to the ground truth, with rotation prediction yielding slightly sharper boundaries in regions of high motion, such as the lung lobes. Image reconstruction, on the other hand, demonstrates better performance in preserving fine-grained textures, particularly in static or slowly moving regions. The difference maps further underscore these trends, with rotation prediction combined with Mini-batch TTT achieving the lowest error in dynamic regions, while image reconstruction paired with Naïve TTT excels in static areas. These qualitative findings align with our quantitative results, demonstrating the robustness of our framework across different self-supervised tasks and adaptation schemes.

\section{Conclusion and Future Works}
\label{sec:conc}
In this study, we introduced a novel TTT framework for 4D medical image interpolation, leveraging self-supervised auxiliary tasks to address domain shifts during inference. Our method significantly improves interpolation accuracy and robustness by adapting the model to new distribution at test time without requiring any labels. Experiments on the Cardiac and 4D-Lung datasets demonstrate the effectiveness of our approach, with consistent improvements in key metrics such as PSNR, SSIM, and NCC, alongside a notable reduction in NMSE. And our framework is highly flexible, supporting multiple self-supervised tasks and adaptation schemes, making it a strong candidate for real-world clinical use.

Despite these advancements, several challenges remain. The computational cost of TTT may limit its practicality in time-sensitive scenarios.   Additionally, while our experiments focused on cardiac and lung datasets, the generalization of the framework to other anatomical structures and imaging modalities remains to be explored. Future work will focus on two key directions: (1) designing more efficient self-supervised tasks that reduce adaptation time while maintaining performance and (2) evaluating its application to broader medical imaging tasks, including segmentation and registration. By addressing these challenges, we aim to further bridge the gap between research and clinical deployment, ultimately enhancing the utility of AI-driven tools in healthcare.

\bibliographystyle{ieeetr} 
\bibliography{references}  

\vspace{12pt}

\end{document}